\documentclass[a4paper,12pt]{article}

\usepackage{epsfig}
\usepackage{graphicx}
\usepackage{amsmath}
\usepackage{multirow}

\newcommand{\be}{\begin{equation}}
\newcommand{\ee}{\end{equation}}
\newcommand{\br}{\begin{eqnarray}}
\newcommand{\er}{\end{eqnarray}}
\newcommand{\ba}{\begin{array}}
\newcommand{\ea}{\end{array}}
\newcommand{\bi}{\begin{itemize}}
\newcommand{\ei}{\end{itemize}}
\newcommand{\bn}{\begin{enumerate}}
\newcommand{\en}{\end{enumerate}}
\newcommand{\bc}{\begin{center}}
\newcommand{\ec}{\end{center}}

\begin{document}

\begin{flushright}
preprint SHEP-06-21\\
\today
\end{flushright}
\vspace*{1.0truecm}

\begin{center}
{\large\bf Another step towards a no-lose theorem \\[0.15cm]
for NMSSM Higgs discovery
at the LHC}\\
\vspace*{1.0truecm}
{\large S. Moretti, S. Munir and P. Poulose}\\
\vspace*{0.5truecm}
{\it High Energy Physics Group, School of Physics \& Astronomy, \\
 University of Southampton, Southampton, SO17 1BJ, UK}
\end{center}

\vspace*{1.0truecm}
\begin{center}
\begin{abstract}
\noindent
We show how the LHC potential to detect a rather light CP-even Higgs boson
of the NMSSM, $H_1$ or $H_2$, decaying into CP-odd Higgs states, $A_1A_1$,
can be improved if Higgs-strahlung off $W$ bosons and (more marginally) off
top-antitop pairs
are employed alongside vector boson fusion as production modes. Our results
should help extracting at least one Higgs boson signal over the
NMSSM parameter space. 
\end{abstract}
\end{center}

\vspace*{3mm}

\noindent
The Minimal Supersymmetric Standard Model
(MSSM) is affected by the so-called `$\mu$-problem'. Its Superpotential 
contains a dimensionful parameter,
$\mu$, that, upon Electro-Weak Symmetry Breaking (EWSB), 
provides a contribution to the masses of both Higgs bosons and Higgsino
fermions. Furthermore, the associated soft Supersymmetry (SUSY) breaking
term mixes the two Higgs doublets. Now,
the presence of $\mu$ in the Superpotential before EWSB 
indicates that its natural value would be
either 0 or the Planck mass $M_P$. On the one hand, $\mu = 0$ would mean that no
mixing is actually generated between Higgs doublets at any scale and the
minimum of the Higgs potential occurs for $< H_d >
= 0$, so that one would have in turn  massless down-type fermions
and leptons after SU(2) symmetry breaking. On the other hand, $\mu \approx M_P$
would reintroduce a `fine-tuning problem' in the MSSM since the Higgs scalars
would acquire a huge contribution $\sim\mu^2$ to their squared masses (thus
spoiling the effects of SUSY, which  effectively removes otherwise quadratically 
divergent contributions to the Higgs mass from SM particles).
Therefore,  the values of this (arbitrary) parameter
$\mu$ are phenomenologically constrained to be close to $M_{\rm SUSY}$
or $M_W$. 

The most elegant solution to the $\mu$-problem is to introduce a new 
singlet scalar field $S$ into the theory and replace the $\mu$-term in 
the MSSM Superpotential by an interaction term\footnote{Hereafter, hatted 
variables describe Superfields
while un-hatted ones stand for the corresponding scalar 
components.}
 $\sim \hat S \hat H_u \hat H_d$. At the same 
time, also the soft term $B\mu H_u H_d$ is replaced by the dimension-4
term $\sim A_{\lambda} S H_u H_d$. When
the extra scalar field $S$ acquires a Vacuum Expectation Value
(VEV), an effective $\mu$ term, naturally of the EW
scale, is generated automatically. 
This idea has been implemented in the Next-to-Minimal
Supersymmetric Standard Model (NMSSM) \cite{NMSSM}, described by
the Superpotential
\begin{equation}\label{WNMSSM}
W_{\rm{NMSSM}}  = {\hat{Q}}{\hat{H}_u}{\bf{h_u}}{\hat{U}^C} 
                + {\hat{H}_d}{\hat{Q}}{\bf{h_d}}{\hat{D}^C} 
                + {\hat{H}_d}{\hat{L}}{\bf{h_e}}{\hat{E}^C}
 + \lambda\hat{S}(\hat{H}_u\hat{H}_d)+\frac{1}{3}\kappa\hat{S}^3,
\end{equation}
where $\hat{S}$ is an extra Higgs iso-singlet Superfield, 
$\lambda$ and $\kappa$ are dimensionless couplings and 
the last ($Z_3$ invariant) term  is required to explicitly break the dangerous
Peccei-Quinn (PQ) U(1) symmetry \cite{PQ}\footnote{One could also gauge the U(1)$_{\rm PQ}$ 
group, so that the $Z_3$ symmetry is embedded in the local gauge symmetry \cite{Z3}.}.  
(See Ref.~\cite{slightly} for NMSSM Higgs sector phenomenology with an exact or
slightly broken PQ symmetry.) However, due to its $Z_3$ symmetry, the NMSSM has a domain wall 
problem, as discussed in the last few references in \cite{DW}. This is to be solved by
additional terms that break $Z_3$ explicitely. Although the latter can
generate dangerous tadpole diagrams, as discussed in the first few references in  \cite{DW}, scenarios that solve both problems 
simultaneously are proposed in \cite{KT}. (Alternative formulations to the NMSSM -- known
as the Minimal Non-minimal Supersymmetric Standard Model (MNSSM) and 
new Minimally-extended Supersymmetric Standard Model or 
nearly-Minimal Supersymmetric Standard Model (nMSSM) --
exist \cite{other-non-minimal}.)
Another positive feature of all these non-minimal SUSY models is that they predict the existence of a (quasi-)stable
singlet-type neutralino (the singlino) that could be responsible for the Dark Matter (DM)
of the universe, albeit this occurs only in limited regions of parameter space \cite{DM}.   
Finally, in these extended SUSY models, the singlet Superfield $\hat{S}$ has no SM gauge group charge
(so that MSSM gauge coupling unification is preserved) and one
can comfortably explain the baryon asymmetry of the Universe
by means of a strong first order EW phase transition \cite{baryon1} (unlike
the MSSM, which requires a light top squark and Higgs boson barely compatible
with experimental bounds \cite{baryon2}).  

Clearly, in eq.~(\ref{WNMSSM}), upon EWSB, a 
 VEV will be generated for the real scalar 
component of $\hat S$ (the singlet Higgs
field), $<S>$, alongside those of the two doublets  $<H_u>$
and $<H_d>$ (related by the parameter $\tan\beta=
<H_u>/<H_d>$). In the absence of fine-tuning, one should expect these
three VEVs to be of the order of $M_{\rm{SUSY}}$ or $M_W$, so that now
one has an `effective $\mu$-parameter', 
$\mu_{\rm{eff}}=\lambda <S>$,
of the required size, thus effectively solving the $\mu$-problem.
In the end, in
the NMSSM, the soft SUSY-breaking Higgs sector is described by the
Lagrangian contribution
\begin{equation}
V_{\rm NMSSM} = m_{H_u}^2|H_u|^2+m_{H_d}^2|H_d|^2+m_{S}^2|S|^2
              + \left(\lambda A_\lambda S H_u H_d + \frac{1}{3}\kappa A_\kappa S^3 + {\rm h.c.}\right),
\end{equation}
with $A_\lambda$ and $A_\kappa$ dimensionful parameters of  ${\cal O}(M_{\rm{SUSY}})$. 

As a result of the introduction of an extra complex singlet scalar
field, which only couples to the two MSSM-type Higgs doublets, the
Higgs sector of the NMSSM comprises of a total of seven mass
eigenstates: a charged pair $H^\pm$, three CP-even Higgses
$H_{1,2,3}$ ($M_{H_1}<M_{H_2}<M_{H_3}$) and two CP-odd
Higgses $A_{1,2}$ ($M_{A_1}<M_{A_2}$). Consequently, Higgs
phenomenology in the NMSSM may plausibly  be different from that of the
MSSM. 


In view of the upcoming CERN Large Hadron Collider (LHC),  
quite some work has been dedicated to probing the Next-to-Minimal
Supersymmetric Standard Model (NMSSM) \cite{NMSSM}
Higgs sector over recent years. Primarily, there have been attempts
to extend the so-called `no-lose theorem' of the MSSM \cite{Dai:1993at} 
to the case of the NMSSM 
\cite{NoLoseNMSSM1a,NoLoseNMSSM1b}\footnote{See
Refs.~\cite{Erice}--\cite{Moretti:2006sv} for a complementary approach, named `more-to-gain theorem',
attempting to define regions of the NMSSM parameter space where  more Higgs
states are visible at the LHC than those available within the MSSM.}. 
From this perspective,
it was realised that at least one NMSSM Higgs boson should remain observable 
at the LHC over the NMSSM parameter space that does not allow any Higgs-to-Higgs 
decay. However, when the only light non-singlet (and, therefore, potentially visible) CP-even
Higgs boson, $H_1$ or $H_2$, decays mainly to two very light 
CP-odd Higgs bosons, $A_1 A_1$, one
may not have a Higgs signal of statistical significance at the LHC. 

From the preliminary studies in Ref.~\cite{NoLoseNMSSM1b} though,
it appeared that using the $qq\to qq W^+W^-,qqZZ\to qq H_{1,2} \to qq A_1A_1$ detection mode, i.e.,
via Vector Boson Fusion (VBF), may lead to the possibility of establishing
a no-lose theorem in the NMSSM, particularly if the lightest CP-odd
Higgs mass is such that there can happen abundant $A_1A_1\to b\bar b\tau^+\tau^-$
decays, with both $\tau$-leptons being detected via
their $e,\mu$ leptonic decays\footnote{The scope of other decays,  $A_1A_1\to jjjj$,
$A_1A_1\to jj\tau^+\tau^-$ (where $j$ represents a 
light quark jet) or $A_1A_1\to  \tau^+\tau^-\tau^+\tau^-$ is very much reduced
in comparison.}. At high luminosity,  this signal may be detectable at the LHC 
as a bump in the tail of a rapidly falling mass distribution. However,
this procedure relies on the background shape to be accurately predictable. 
These analyses were based on Monte Carlo
(MC) event generation (chiefly, via the SUSY routines
of the {\tt HERWIG} v6.4 code \cite{SHERWIG}) and a toy detector simulation ({\tt GETJET}, 
based on UA1 software). Further analyses based on {\tt PYTHIA} v6.2
\cite{Sjostrand:2001yu} and a more proper ATLAS detector 
simulation  ({\tt ATLFAST}) \cite{Baffioni:2004gdr} found that the original selection
procedures may need improvement in order to extract a signal \cite{Orsay}. 

While the jury is still out on this particular analysis, we would like here to
advertise the possibilities offered by exploiting Higgs-strahlung (HS) off
gauge bosons
($q\bar q'\to W^{\pm*}\to W^\pm H_{1,2}$, with a subleading component
from $q\bar q\to Z^{0*}\to Z^0 H_{1,2}$) and, more marginally, off
heavy quark pairs (chiefly top quarks, $q\bar q,gg\to t\bar t H$, because
of the small $\tan\beta$ values involved in the scenarios outlined
in \cite{NoLoseNMSSM1b})
as the underlying Higgs production modes, in place of or -- better -- alongside VBF. In fact,
for the $H_{1,2}$ masses of relevance to the above analyses, say, 50
to 120 GeV, Higgs-strahlung gives cross sections comparable
to VBF, if not larger for smaller $M_{H_{1,2}}$ values. However, we will not be performing here 
a detector analysis, including parton shower and hadronisation effects, as
in \cite{NoLoseNMSSM1b,Baffioni:2004gdr}.
Rather, in this brief report,  we will limit ourselves to proving that, after enforcing standard 
LHC triggers (at partonic level)
 on $W$ decays in Higgs-strahlung and on 
forward/backward jets in VBF, there are regions of NMSSM
parameter space were the yield of the former is of the same size as that of the latter,
no matter what the $A_{1}A_1$ decay pattern may be. Therefore, we conclude that our results
are encouraging in an attempt to establish the aforementioned NMSSM no-lose theorem at the LHC.

For a general study of the NMSSM Higgs sector (without any assumption on the
underlying SUSY-breaking mechanism) we used here the 
{\tt NMHDECAY} code  (version 1.1) \cite{Ellwanger:2004xm}. 
(We have verified that the pattern described below does not change 
if one adopts the newest version \cite{NMHDECAY2}.)
This program
computes the masses, couplings and decay Branching Ratios (BRs) of all NMSSM Higgs
bosons in terms of the model parameters taken at the EW
scale. The
computation of the spectrum includes leading two-loop terms,
EW corrections and propagator corrections.  {\tt NMHDECAY} 
also takes into account theoretical as well as
experimental constraints from negative Higgs searches at 
collider experiments. For our purpose, instead of postulating unification, we fixed the soft SUSY breaking terms to a very high value, 
so that they have little or no contribution to the outputs of the parameter scans. 
Consequently, we are left with six free parameters: the usual tan$\beta$, the Yukawa 
couplings $\lambda$ and $\kappa$, the soft trilinear terms $A_\lambda$ and $A_\kappa$ plus 
$\mu_{\rm eff} = \lambda\langle S\rangle$. 

We have used {\tt NMHDECAY} to scan over the NMSSM
parameter space defined in \cite{Moretti:2006sv} (borrowed from 
\cite{Ellwanger:2005uu}), with the aforementioned six parameters 
taken in the following intervals\footnote{Notice that a top quark pole mass of 
$m_t=175$ GeV was used as default, though we have verified that values within
current error bands (see \cite{topmass}) 
have a numerically small impact on our analysis, thus
leaving the main conclusions of the paper unchanged.}:
\begin{center}
$\lambda$ : 0.0001 -- 0.75,\phantom{aa} $\kappa$ : $-$0.65 --
+0.65,\phantom{aa} $\tan\beta$ : 1.6 -- 54,\\ $\mu$, $A_{\lambda}$,
$A_{\kappa}$ :  $-$1000 -- +1000 GeV.\\
\end{center}
Remaining soft terms which are fixed in the scan include:\\
$\bullet\phantom{a}m_{Q_3} = m_{U_3} = m_{D_3} = m_{L_3} = m_{E_3} = 2$ TeV, \\
$\bullet\phantom{a}A_{U_3} = A_{D_3} = A_{E_3} = 1.5$ TeV,\\
$\bullet\phantom{a}m_Q = m_U = m_D = m_L = m_E = 2$ TeV,\\
$\bullet\phantom{a} M_1 = M_2 = M_3 = 3$ TeV.\\

The allowed decay  modes for neutral NMSSM Higgs bosons are
into any SM particle, plus into any final state
involving all possible combinations
 of two Higgs bosons (neutral and/or charged)
or of one Higgs boson and a gauge vector as well as into
all possible sparticles.
We have performed our scan over several millions of randomly
selected points in the specified parameter space.  
The data points surviving  all constraints are then used to determine the
cross sections for NMSSM Higgs hadro-production. As the SUSY mass scales 
have been set well above the EW one, 
the production modes
exploitable in simulations at the LHC are the usual ones, 
 the so-called `direct' Higgs production modes
of \cite{Kunszt:1996yp}.

\begin{figure}[h!]
\hspace*{3.5truecm}{\epsfig{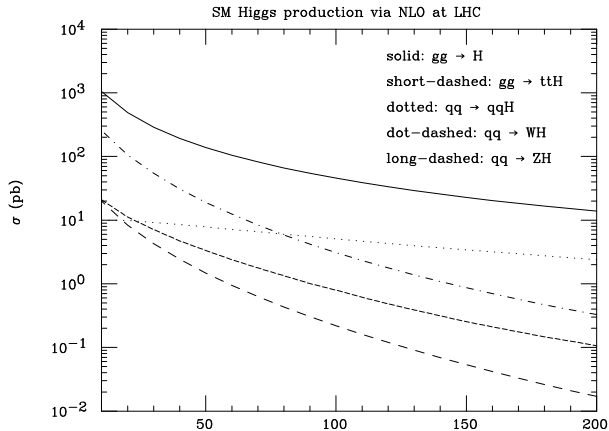}}
\caption{The Higgs production cross sections through NLO QCD in the SM at the LHC.
}
\label{fig:XsectSM}
\end{figure}

As we are aiming at comparing the yield of VBF ({\tt $qq\to qqH$}) against 
HS off $W$ bosons (W-HS)
({\tt $qq\to WH$}) and off $t\bar t$ pairs (tt-HS) ({\tt $gg\to ttH$}), it is of relevance to study in Fig.~\ref{fig:XsectSM}
the light Higgs, $H$, hadro-production cross sections at the
LHC in the SM, as the NMSSM rates 
would be obtained from these (for a given Higgs
mass) by rescaling the $VVH$ and $ttH$ couplings. 
We see that in the SM W-HS dominates for Higgs masses below
80 GeV while VBF becomes the leading channel above such a value
(in the NMSSM these two processes are rescaled by the same amount). The case
tt-HS is generally subleading (even in presence of appropriate NMSSM
couplings), but not negligible at low Higgs masses.
Besides, as intimated earlier, notice that HS off $Z$ boson is always 
very small, so we will ignore it in the remainder
of the paper. It is also
worth recalling that gluon-gluon fusion ({\tt $gg\to H$}), despite being
the mode with largest production rates, plays no role in our case, as $H_{1,2}\to
A_1A_1$ decay channels would not be extractable in this case from the background.
 (Notice in the figure the normalisation via NLO QCD throughout.) 

\begin{figure}
\begin{tabular}{cc}
\hspace*{1.truecm}
\includegraphics[scale=0.505]{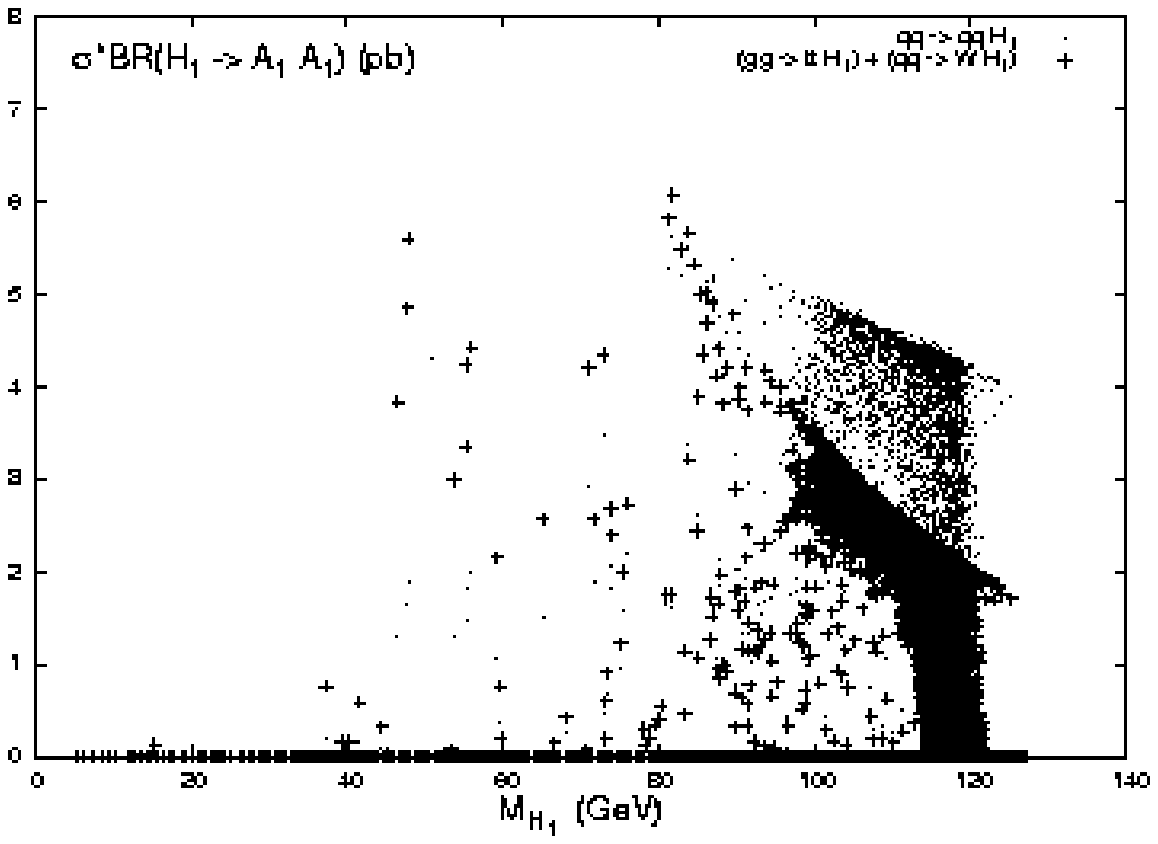}&
\includegraphics[scale=0.505]{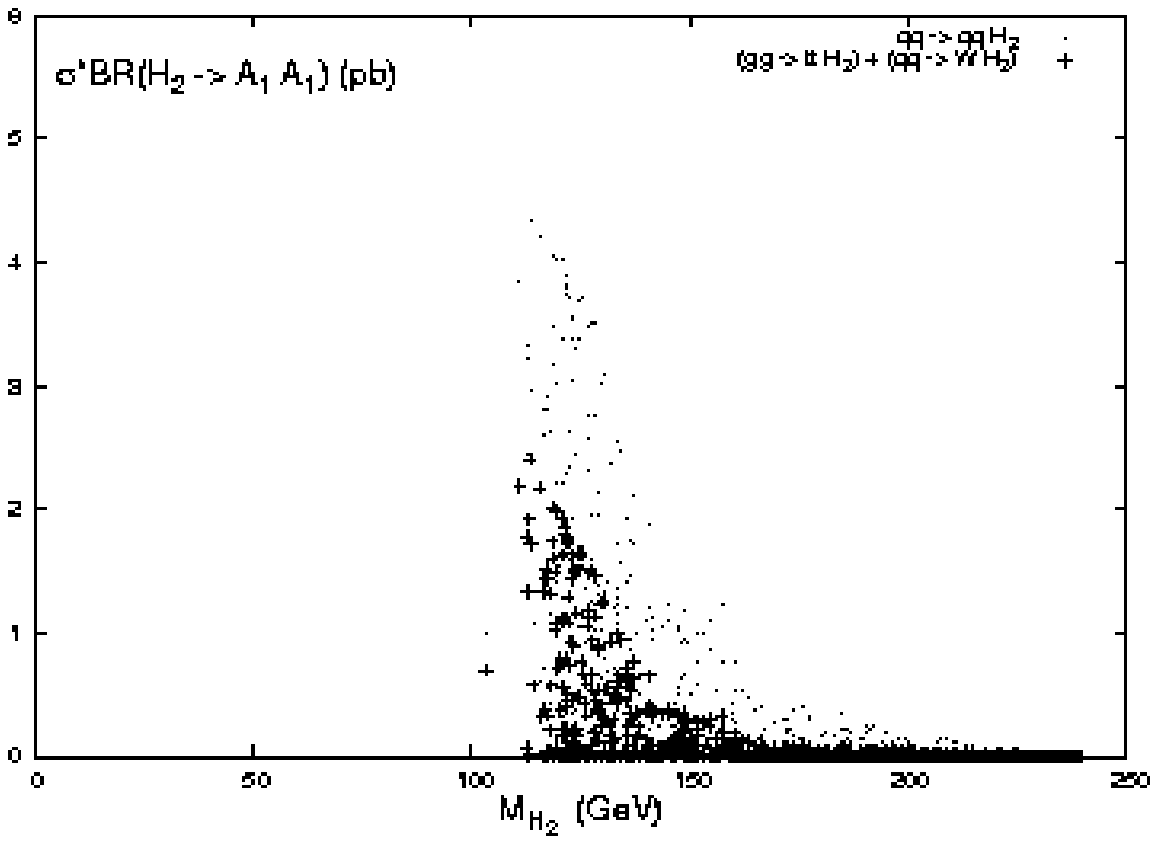}
\end{tabular}
\caption{Cross section times BR of $H_1$ (left) and $H_2$ (right) 
plotted against their respective masses. The symbol `$\cdot$' refers to VBF while `$+$' to W-HS + tt-HS.}
\label{fig:XsectBRH}
\end{figure}

As a second step we computed the NMSSM total cross section
times BR into $A_1A_1$ pairs for VBF and W-HS + tt-HS  for each of the two
lightest neutral Higgs bosons, $H_1$ and $H_2$. We display
these rates in Fig.~\ref{fig:XsectBRH}
as a function of $M_{H_1}$ and $M_{H_2}$. Here,
one can appreciate that there exist more possibilities of
establishing a $H_1$ signal than one due to $H_2$. Whereas
the potential to detect the heavier of these two
Higgs states is confined to masses above 115
GeV or so and probably below 140 GeV, 
where VBF is largely dominant with respect to W-HS + 
tt-HS,
in the case of the light state  
there exists a low mass window where production rates via
the latter two processes combined are comparable to those from the former, 
most often within 10--20\% from each other. In fact, at times, W-HS + tt-HS
rates are larger than those for VBF, the more so the lower the
$H_1$ mass. (Recall that all parameter points examined here are
compliant with collider bounds, even those at very low Higgs mass,
as these correspond to reduced Higgs couplings to gauge bosons.)
Now, one should bear in 
mind that the rates in Fig.~\ref{fig:XsectBRH}
do not include yet the efficiency to trigger on the signal.
In the case of VBF, one triggers on one forward and one backward
jet, with  $p_T    >  20$    GeV,
 $|\eta|    <  5$ and
 $\eta({\rm fwd})\cdot\eta({\rm bwd})    < 0$. The efficiency is here about 60\%.
In the case of W-HS, one triggers on a high transverse momentum
lepton (electron or muon), with  $p_T    >  20$    GeV and
 $|\eta|    <  2.5$. In this case the efficiency is lower, about
19\%, primarily due to the fact that a $W$ boson decays into 
electron/muons only about 20\% of the times. {The efficiency for 
tt-HS is 14\%, as one top is required
to decay hadronically and the other leptonically.}
(Note that the efficiency values quoted are basically independent of the
Higgs mass.) Even so, the W-HS component, aided by the tt-HS one, would make a sizable addition
to the production rates of VBF. As we expect the efficiency of extracting
whichever $H_{1,2}\to A_1A_1$ decays to be the same in both 
processes\footnote{If anything, since no actual $b$-tagging was enforced in
the analyses of Refs.~\cite{NoLoseNMSSM1b,Baffioni:2004gdr},
whenever $A_1A_1$
hadronic decays are present, we would expect the efficiency to
worsen for the case of VBF, because of jet combinatorics. }, we see a potential
in improving the signal yield by using all mentioned channels, 
beyond what achieved by using VBF alone. 

By recalling that the efficiency
to trigger on VBF is at least three times the one to isolate W-HS + tt-HS, it is of particular
interest to estimate the proportion of points where the latter gives more cross
section than the former. Despite we found that W-HS + tt-HS very rarely exceeds VBF by more 
than a factor of three, there are clear zones of NMSSM parameters space where 
W-HS + tt-HS is consistently larger than VBF, those producing $M_{H_1}$ values below
80 GeV, indeed the SM crossing point seen in Fig.~\ref{fig:XsectSM}. 
Evidently, this mass range is of relevance to $H_1\to A_1A_1$ decays
only, see Fig.~\ref{fig:XsectBRH}. In fact, for the case of $H_2\to A_1A_1$, cross sections  
are much smaller in comparison and VBF is always
very dominant, as -- for potentially detectable rates -- $M_{H_2}$ is
above $\approx115$ GeV and below $\approx140$ GeV. Finally, notice that $H_2\to H_1H_1$ decays 
very often compete
with $H_2\to A_1A_1$ \cite{Ellwanger:2005uu}. In fact the former occur almost as often
as the latter over the NMSSM parameter space investigated here. To make use of this
channel too, a slight modification of the procedures advocated in \cite{NoLoseNMSSM1b}
would be required. 

\begin{figure}
\begin{tabular}{cc}
\hspace*{1.0truecm}\includegraphics[scale=0.505]{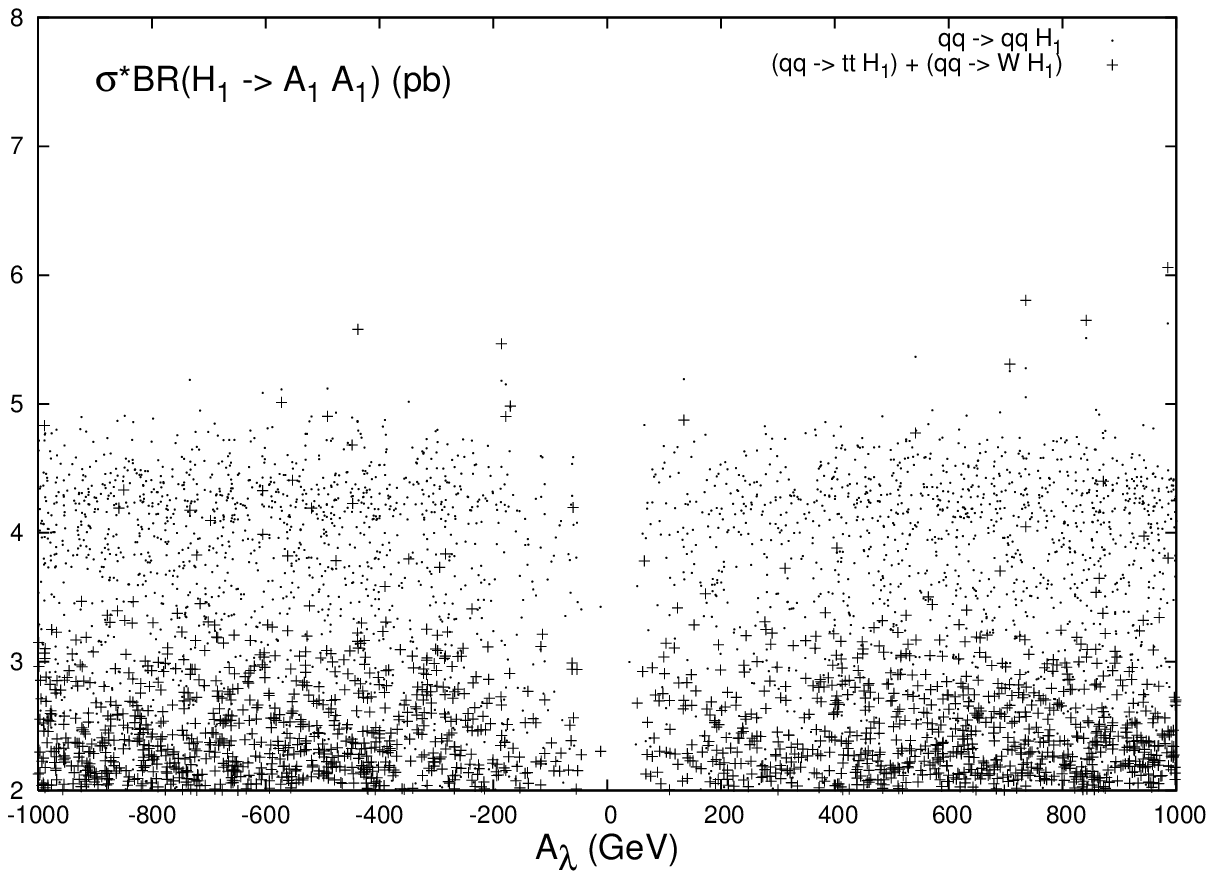}&
\hspace*{1.0truecm}\includegraphics[scale=0.505]{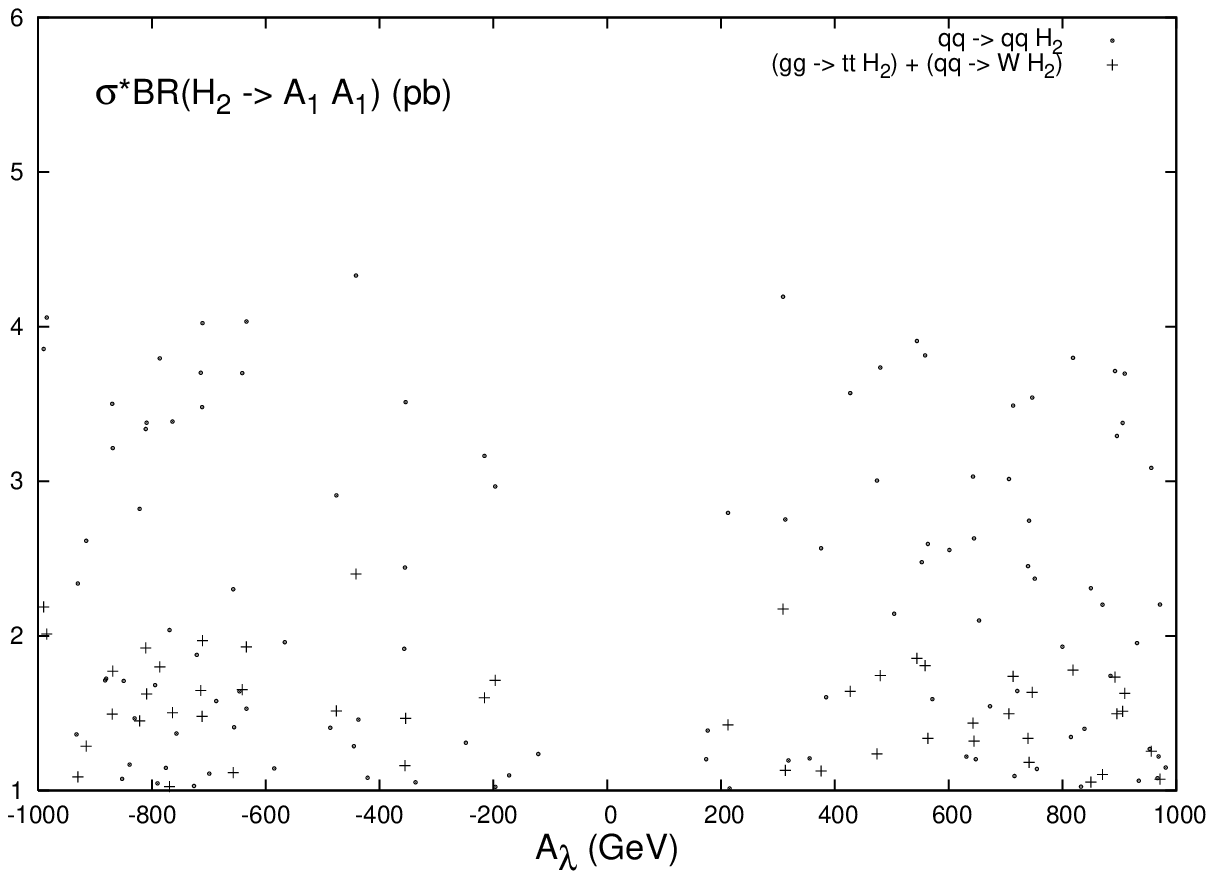}\\
\hspace*{1.0truecm}\includegraphics[scale=0.505]{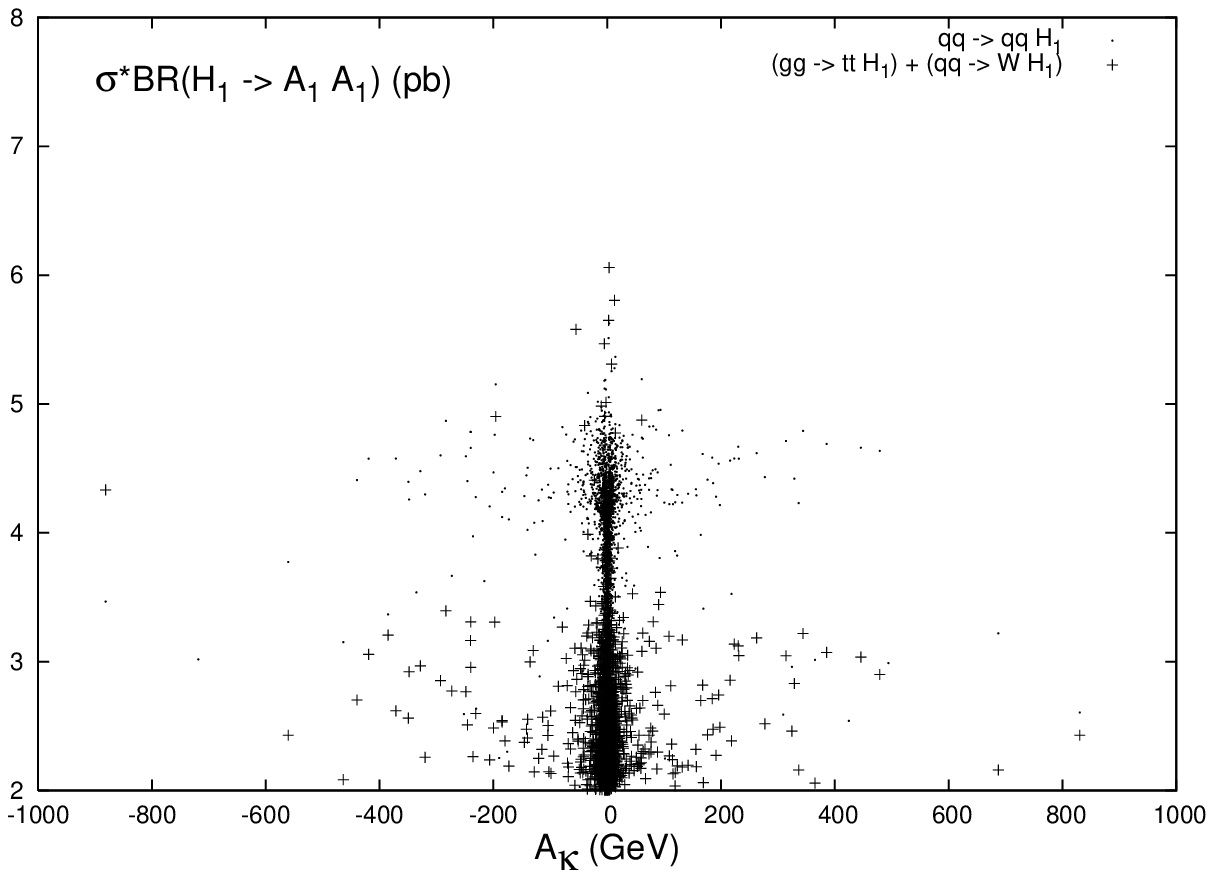}&
\hspace*{1.0truecm}\includegraphics[scale=0.505]{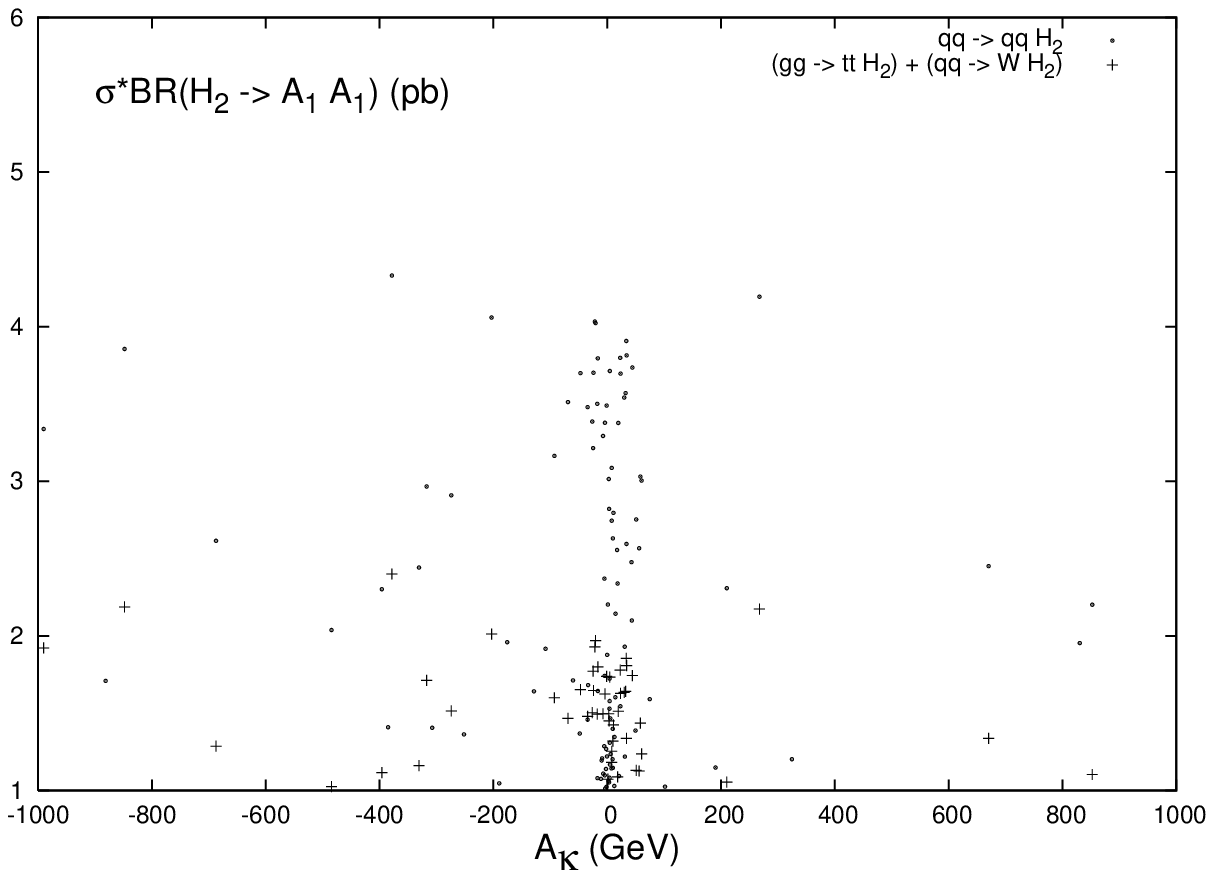}\\
\hspace*{1.0truecm}\includegraphics[scale=0.505]{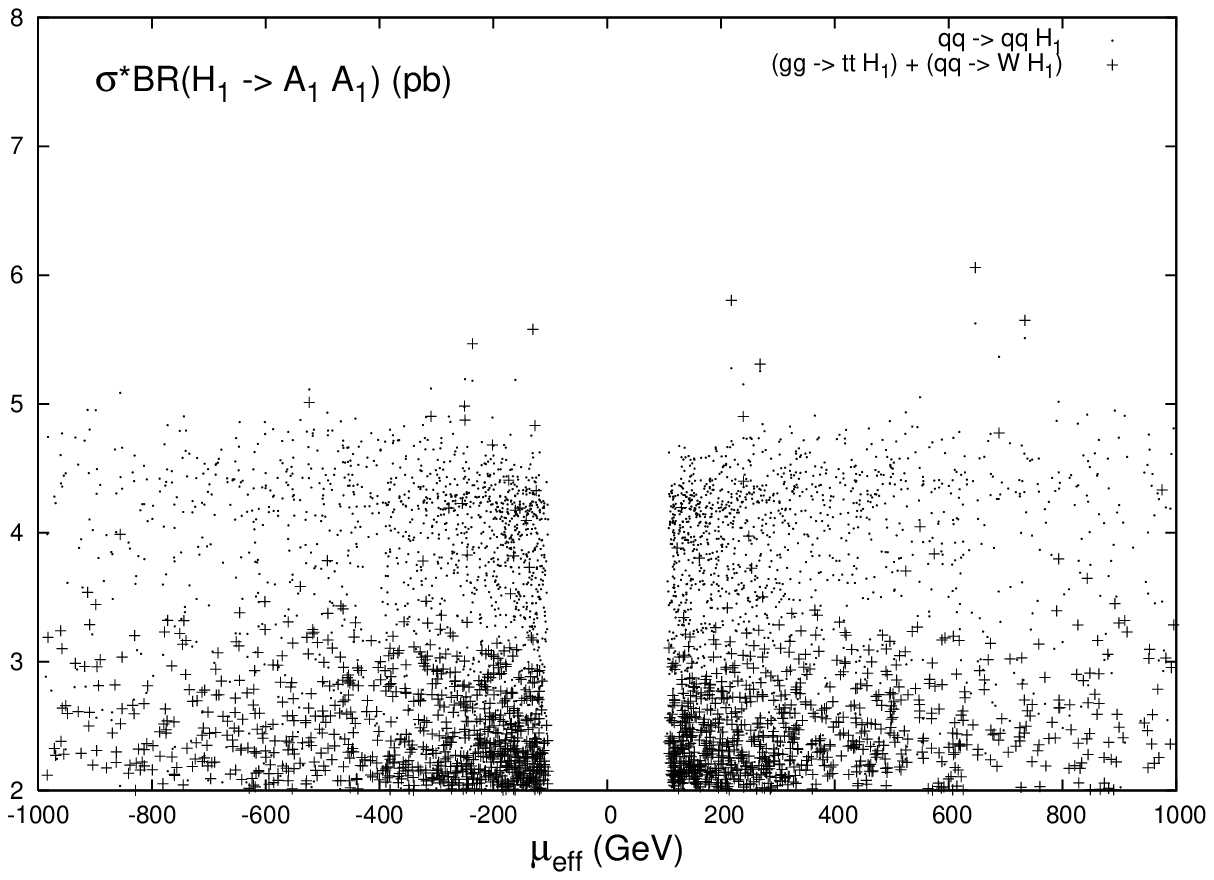}&
\hspace*{1.0truecm}\includegraphics[scale=0.505]{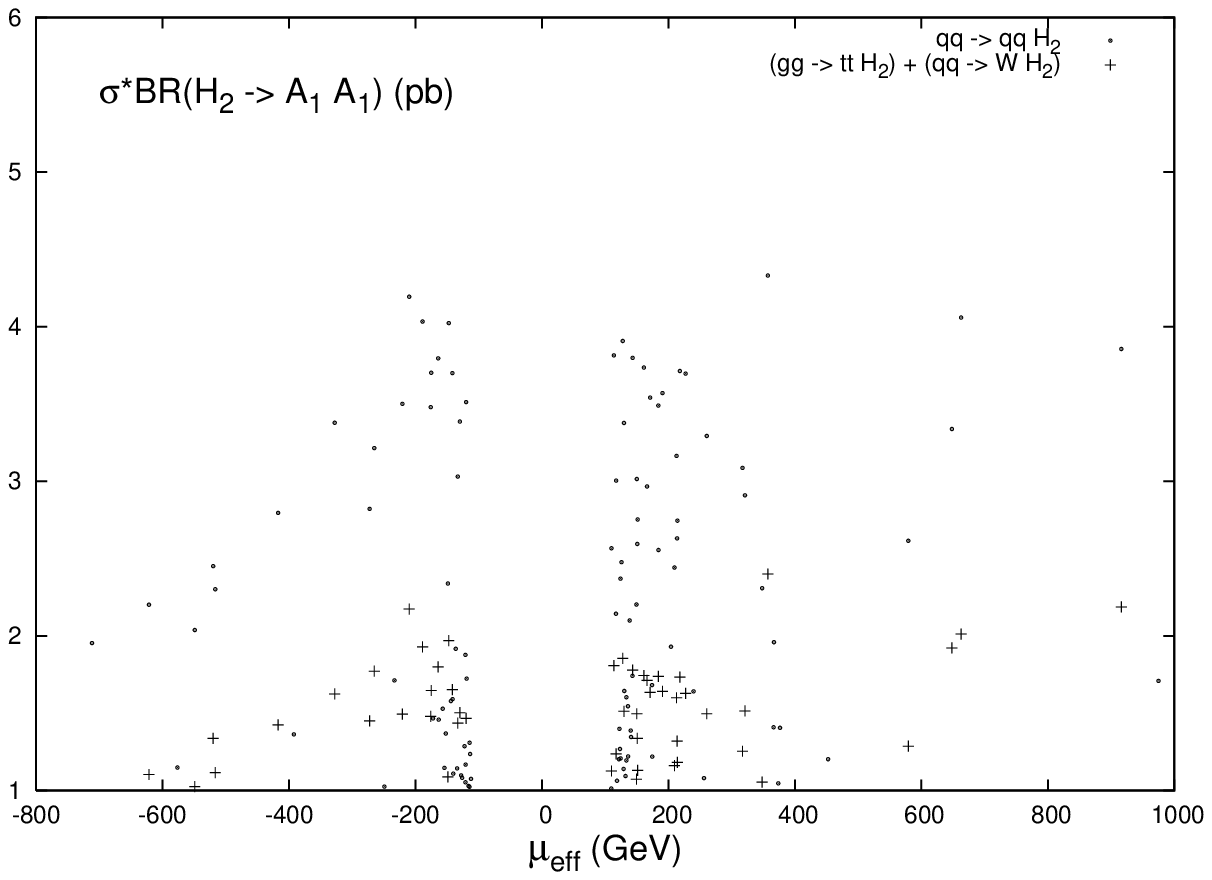}
\end{tabular}
\vspace*{1.0truecm}\centerline{(b)}
\caption{Cross section times BR of $H_1$ (left) and $H_2$ (right) when potentially visible, i.e., limited to those NMSSM parameter
points for which both cross sections times BRs are larger than 2(1) pb for $H_1(H_2)$, plotted against 
the following parameters: (a)
$\tan\beta$, $\lambda$, $\kappa$; (b) $A_\lambda$, $A_\kappa$ and  $\mu_{\rm{eff}}$.}
\label{fig:paramsH}
\end{figure}

Even after accounting
for the trigger efficiencies, the VBF cross sections plotted in
Fig.~\ref{fig:XsectBRH} are in the same range as
those probed in \cite{NoLoseNMSSM1b}\footnote{We have in fact been able
to reproduce most of the points discussed therein.}, so that, for similar
$M_{H_1}$ and $M_{H_2}$ masses, we would expect to obtain the same
overall detection efficiencies seen back then also for all our points 
falling in the mass range, say, 50 to 120 GeV. Crucially,  NMSSM parameter points giving the highest cross sections
for VBF are the same yielding the largest rates for W-HS + tt-HS.
More in general, from Figs.~\ref{fig:paramsH}a--b, one can also gather
where the regions of highest cross sections, for both channels (VBF and W-HS + tt-HS)
and Higgs flavours ($H_1$ and $H_2$), lie in the NMSSM parameter space. In particular,
their distribution is quite homogeneous as they are not located in some specific
areas (i.e., in a sense, not `fine-tuned'). Altogether, the proportion of parameter
space where the two production modes yield potentially detectable Higgs signals
(at least according to the analysis in \cite{NoLoseNMSSM1b}), say, above 1--2 pb
(prior to including tagging efficiencies and $A_1$ decay rates), is 0.21\% for
VBF and 0.13\% for W-HS + tt-HS. However, if production cross sections of 4 pb or upwards
are required to render the $H_{1}\to A_1A_1$ signal visible, then the rates
reduce to $0.096\%$ and $0.0019\%$, respectively. For the case of  $H_{2}\to A_1A_1$,
the numbers are typically 20 and 10 times smaller, for the case of VBF and W-HS + tt-HS,
respectively. 

Clearly, while the production cross sections (after triggering), the selection
procedures and efficiencies to extract the Higgs decays may well be the same in 
both samples, the background will differ. In fact, whilst in the case of VBF the latter
is dominated by top-antitop pair production and 
decay for V-HS and tt-HS we expect that (more manageable)
$WZ$~+~jets events will be the largest noise,
assuming the most promising Higgs signature discussed above (i.e., $b\bar b\tau^+\tau^-$).
A detailed phenomenological study, based
upon parton shower, hadronisation and detector simulation (like  
in Refs.~\cite{NoLoseNMSSM1b,Baffioni:2004gdr}), is obviously in order before
drawing any firm conclusions from our very preliminary study. 
(In this respect, it is also interesting to see how the mass of the decaying Higgs bosons,
$H_1$ and $H_2$, relates to that of the light $A_1$ state: this is illustrated
in Fig.~\ref{fig:mass}.) Nonetheless, we thought
it worthwhile to alert the LHC experiments to the possibility of supplementing
the search for $H_{1,2}\to A_1A_1$ signals via VBF with that through W-HS + tt-HS, as such
Higgs decays are relevant in a region of NMSSM parameter space where the two
production modes are competitive. Whilst the efficiency of tagging
two forward/backward jets in VBF is three times higher than that
to trigger on a high transverse momentum electron/muon in W-HS + tt-HS
(mainly in virtue of the leptonic BR suppression in the second case),
the combination of the latter two remains competitive with the former
 over the Higgs mass range relevant to these
decays, 50 to 120 GeV or so,  the more so the lighter the mass of the 
decaying Higgs state. (Notice that such a low mass scenario is one
alleviating  the so-called
`little fine-tuning problem' of the MSSM, resulting in LEP failing
to detect a light CP-even Higgs boson, predicted over most
of the MSSM parameter space, as in the NMSSM
the mixing among more numerous CP-even or CP-odd
Higgs fields enables light mass states being produced at LEP
yet they can remain undetected because of their reduced couplings
to $Z$ bosons.) Thus, the chances 
of establishing a no-lose theorem in the NMSSM at the LHC via
the aforementioned Higgs-to-Higgs decay mode might improve considerably
if the Higgs state strongly coupled to gauge bosons is the lightest one.
Our analysis was based on a fairly extensive scan of the NMSSM parameter
space incorporating the latest experimental constraints. Detailed MC event generation
studies will be available soon.

\subsection*{Acknowledgements} 
\noindent
SM thanks Cyril Hugonie for discussions. PP's research
is supported by the Framework Programme 6 via a Marie Curie International Incoming Fellowship,
contract number MIF1-CT-2004-002989.

\clearpage
\thispagestyle{empty}
\begin{figure}
\begin{tabular}{cc}
\hspace*{1.0truecm}\includegraphics[scale=0.505]{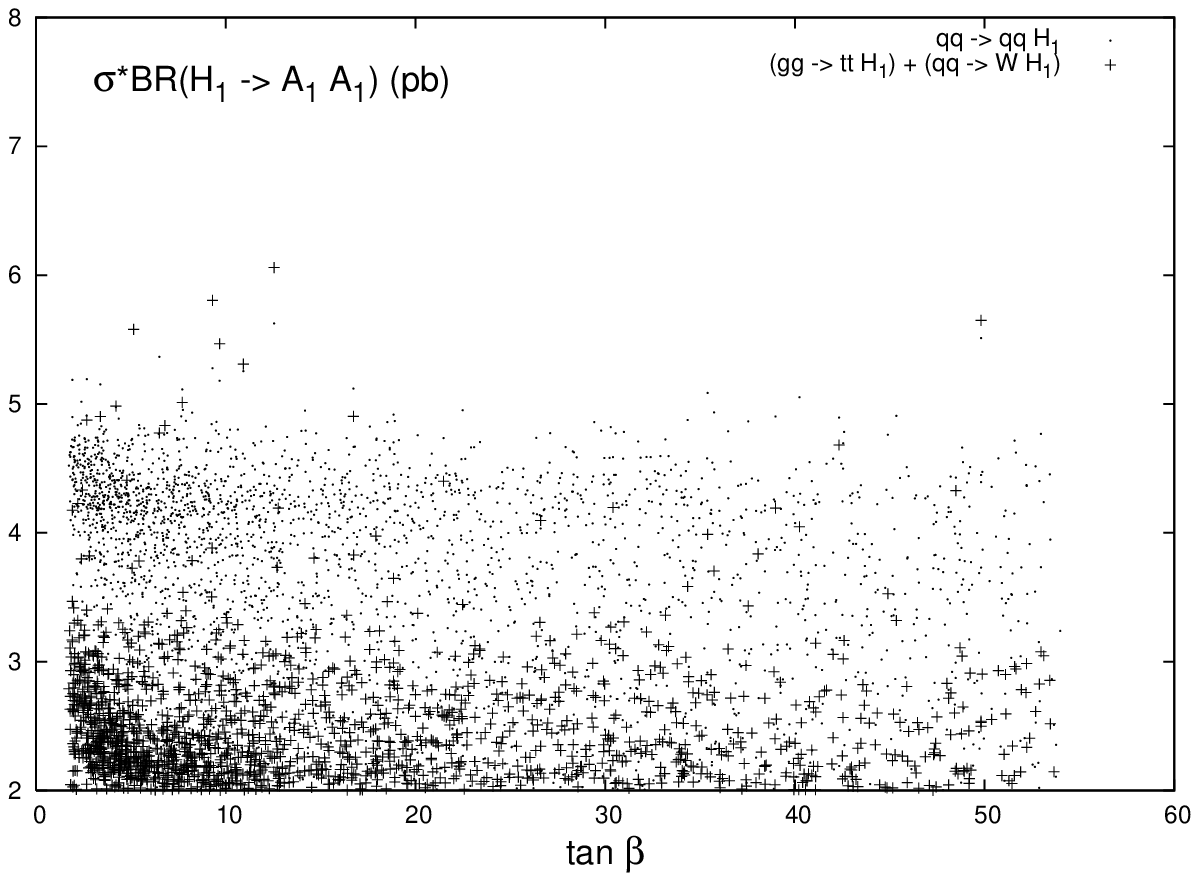}&
\hspace*{1.0truecm}\includegraphics[scale=0.505]{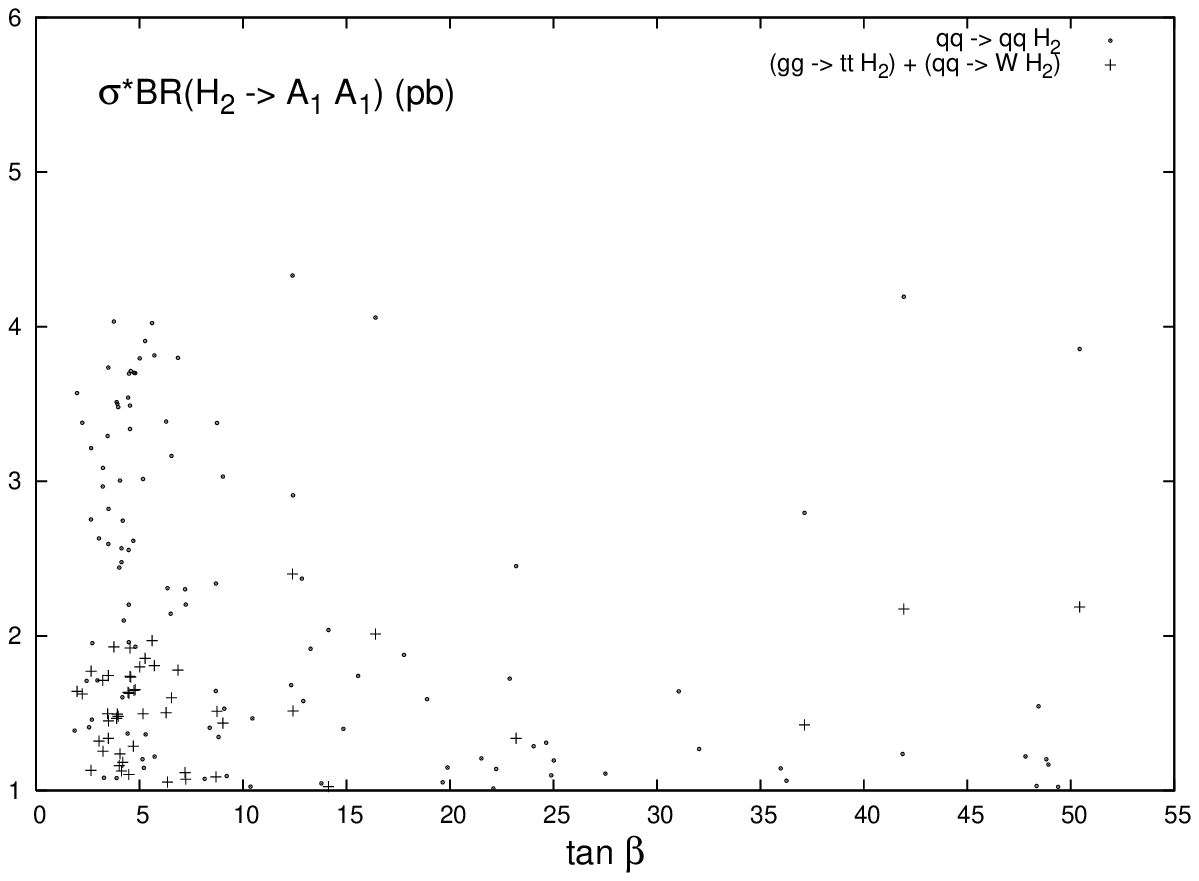}\\
\hspace*{1.0truecm}\includegraphics[scale=0.505]{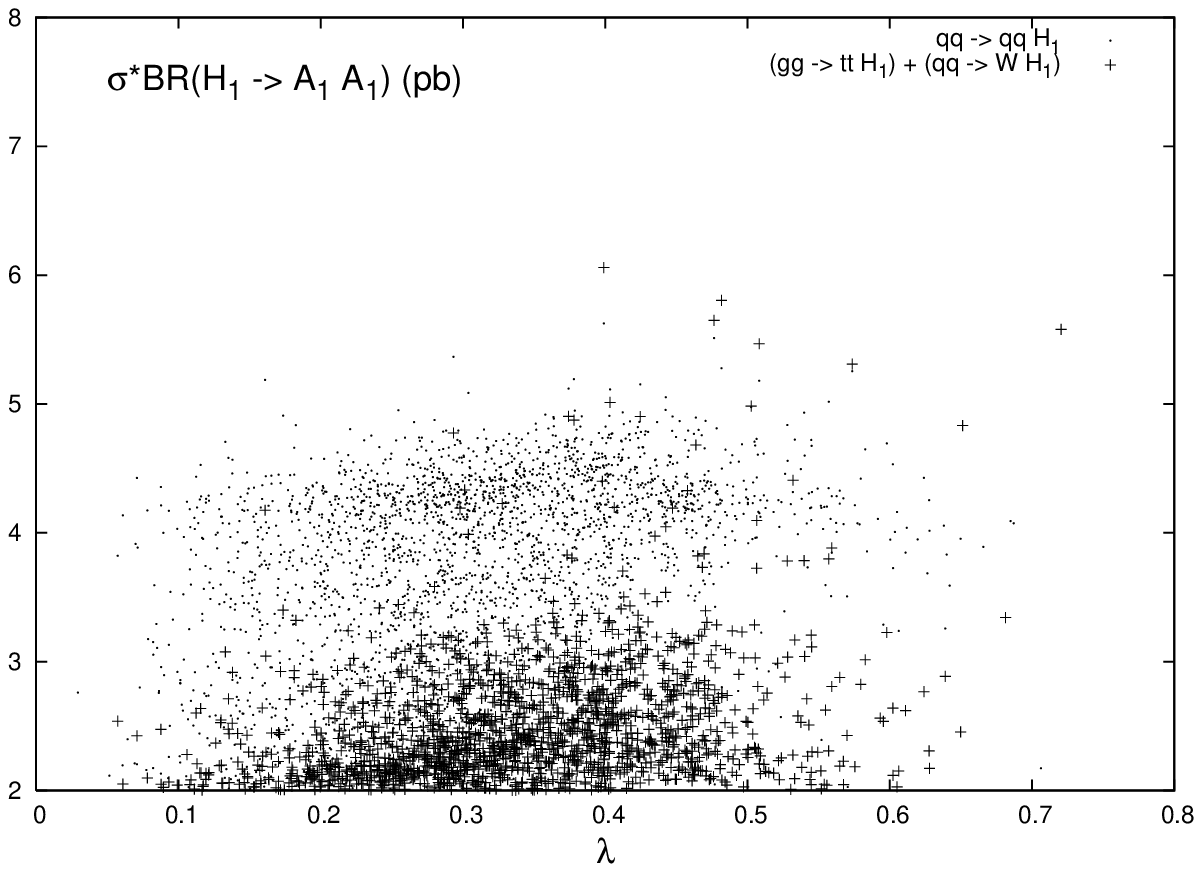}&
\hspace*{1.0truecm}\includegraphics[scale=0.505]{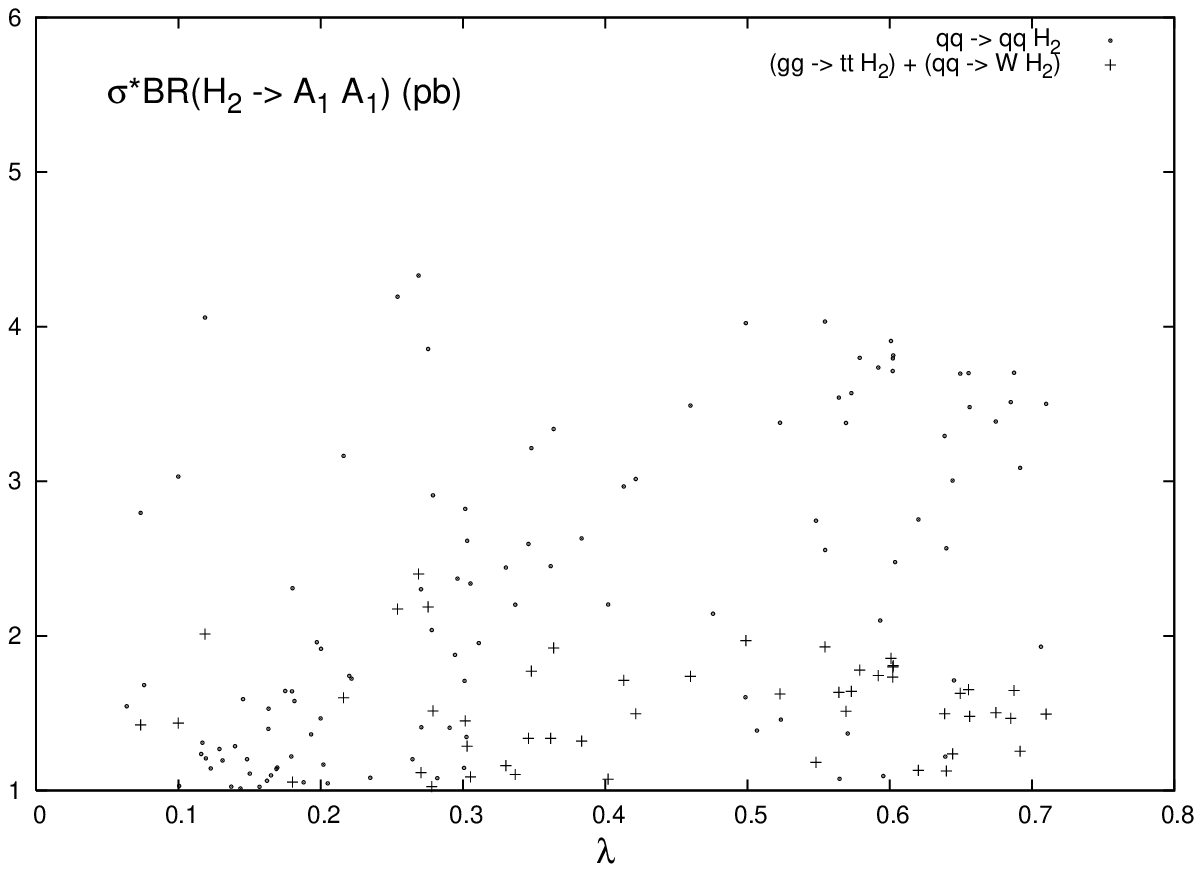}\\
\hspace*{1.0truecm}\includegraphics[scale=0.505]{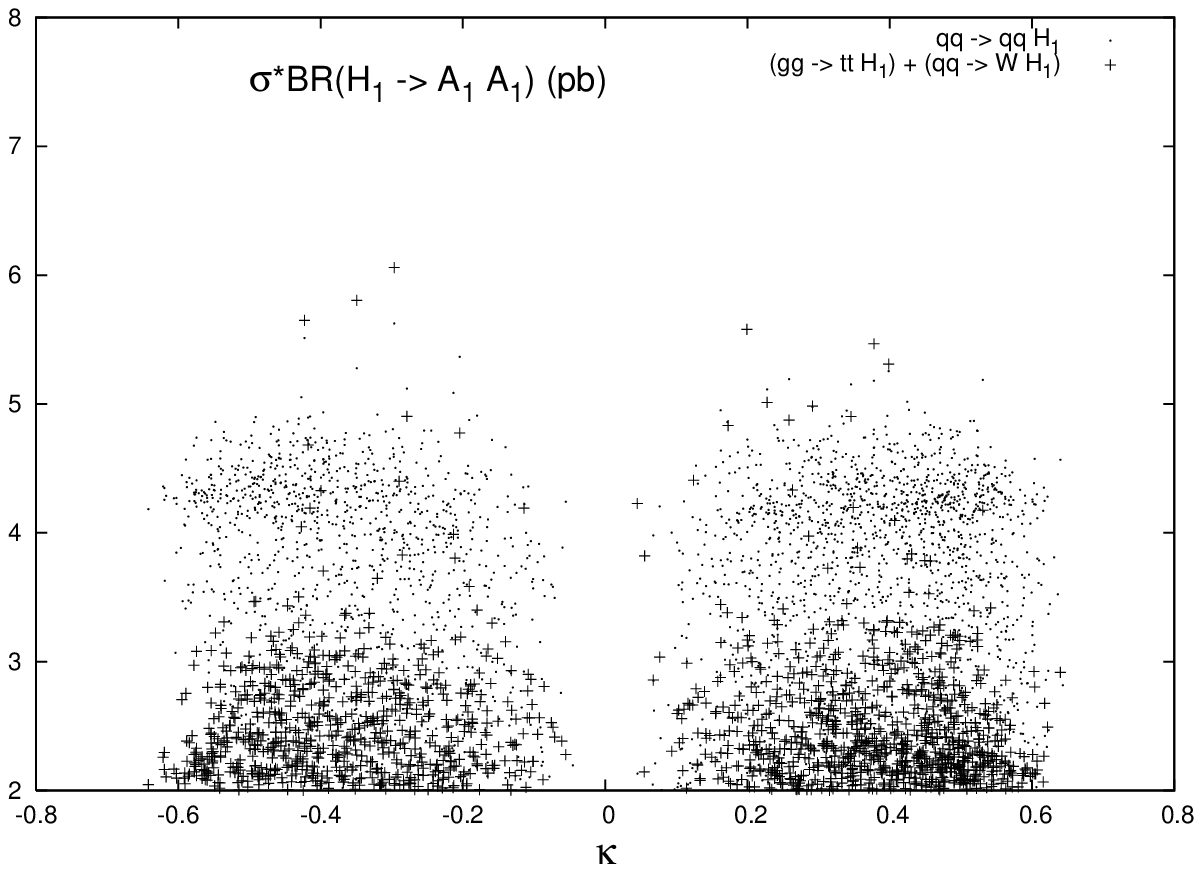}&
\hspace*{1.0truecm}\includegraphics[scale=0.505]{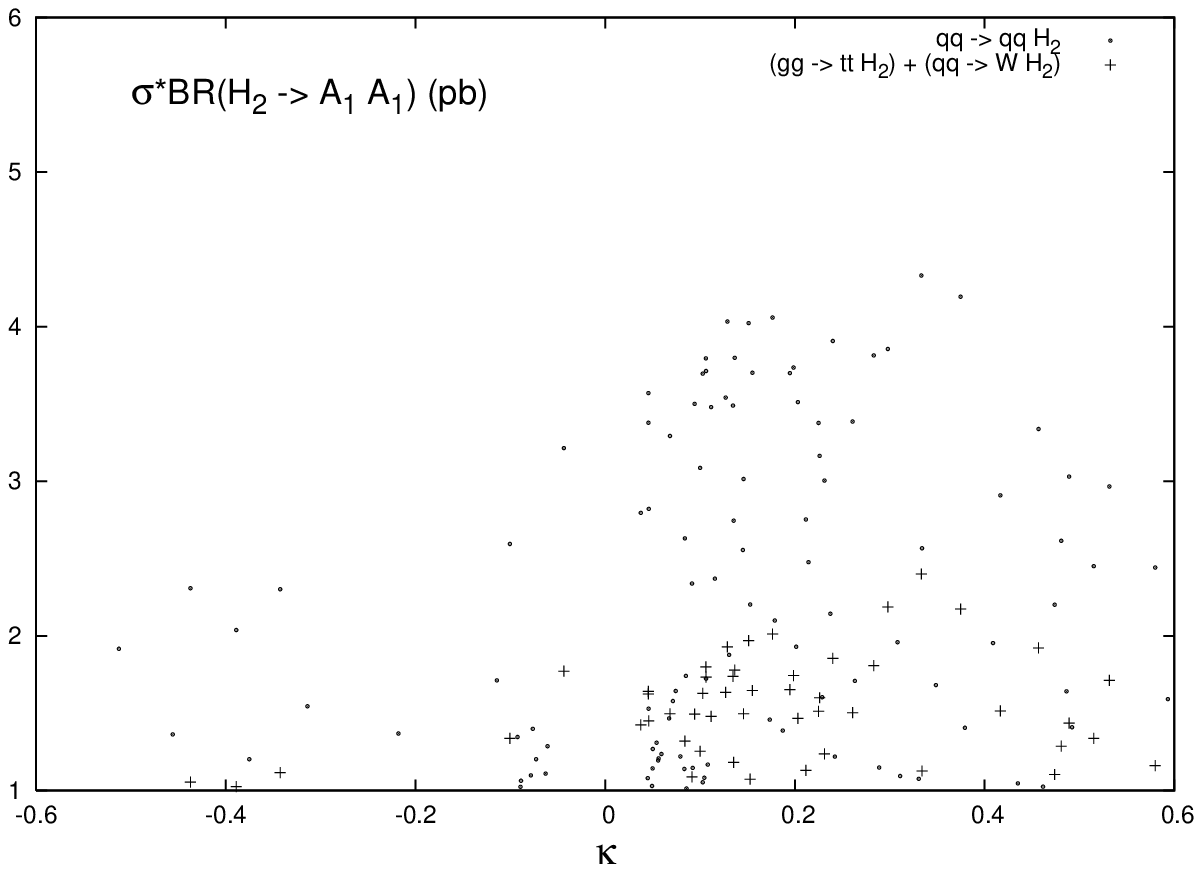}\\
\end{tabular}
\vspace*{1.0truecm}\centerline{(a)}
\end{figure}

\clearpage
\thispagestyle{empty}

\clearpage
\thispagestyle{empty}
\begin{figure}
\begin{tabular}{cc}
\hspace*{1.0truecm}\includegraphics[scale=0.505]{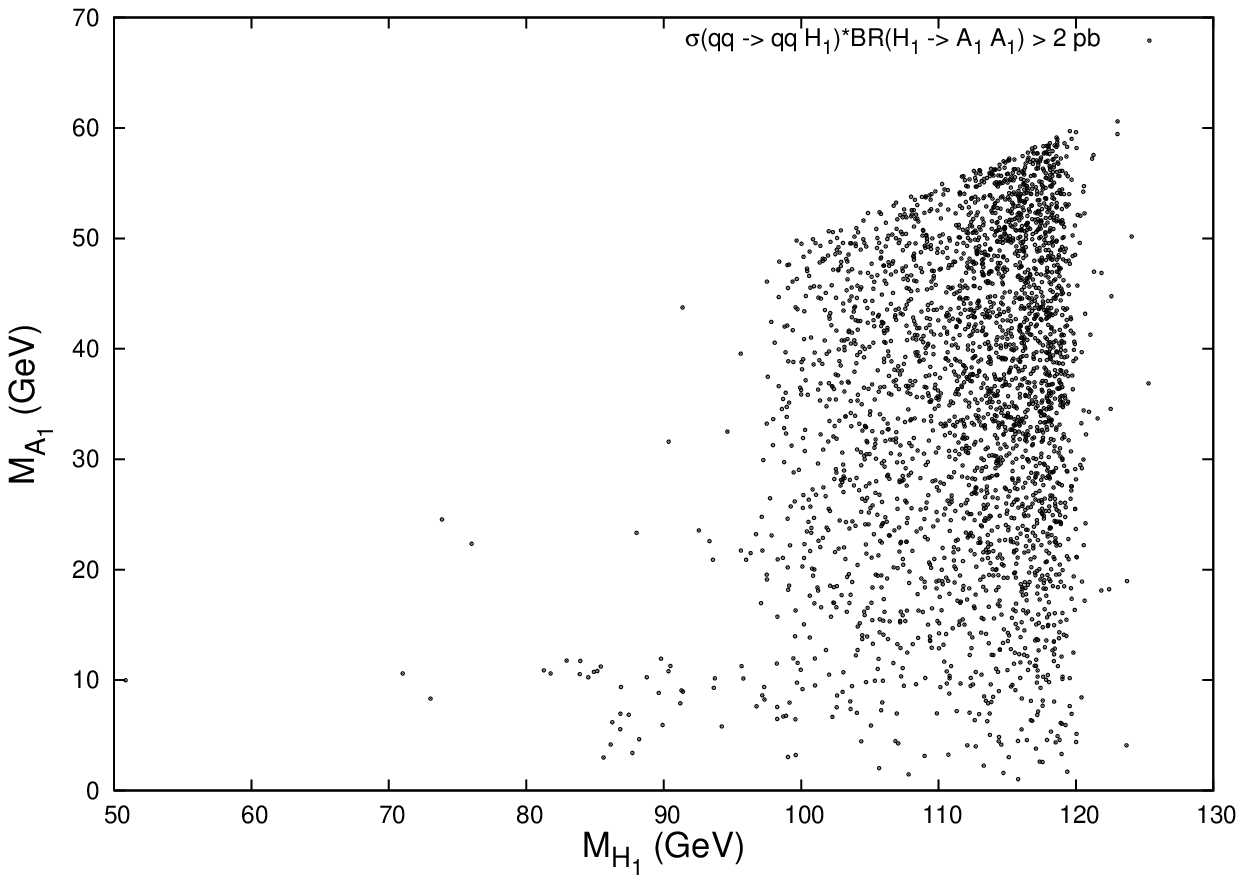}&
\hspace*{1.0truecm}\includegraphics[scale=0.505]{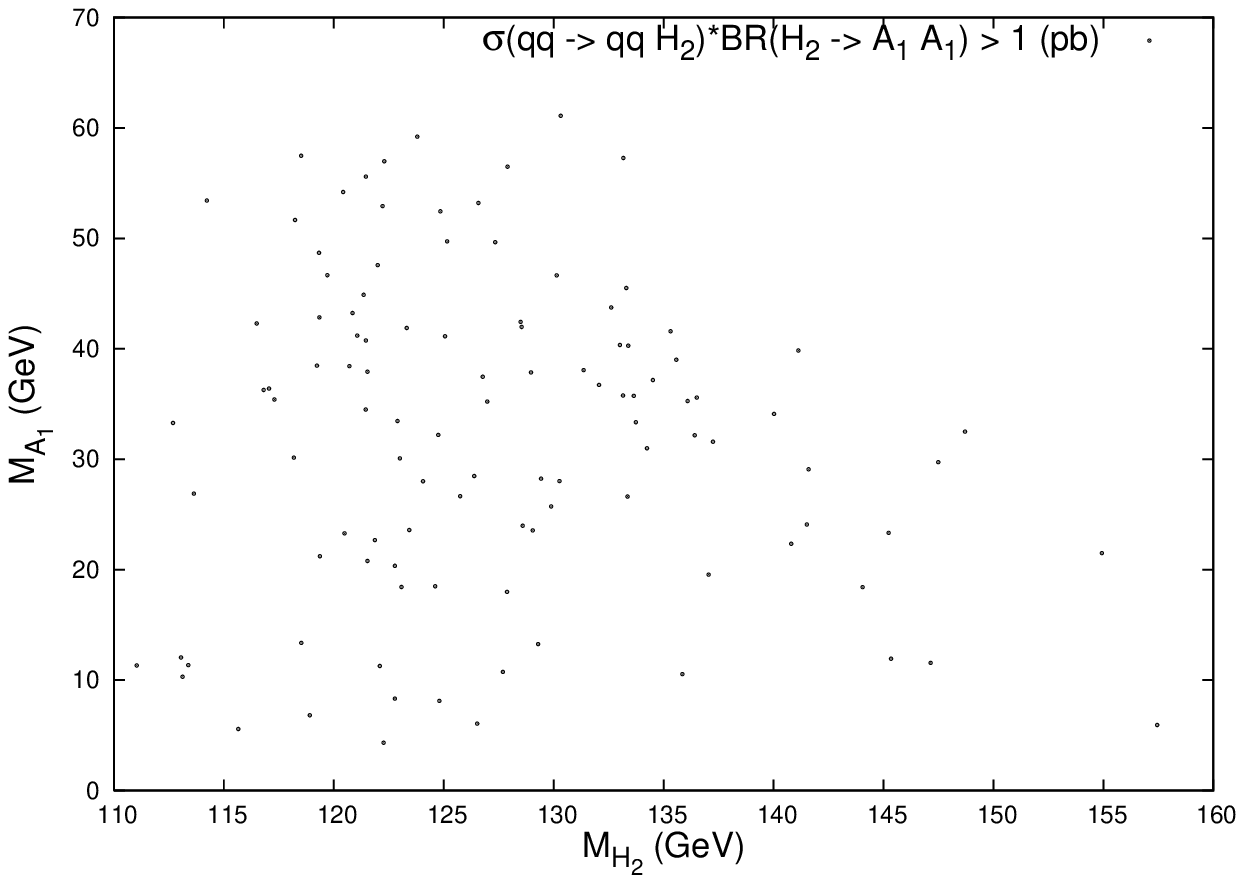}\\
\hspace*{1.0truecm}\includegraphics[scale=0.505]{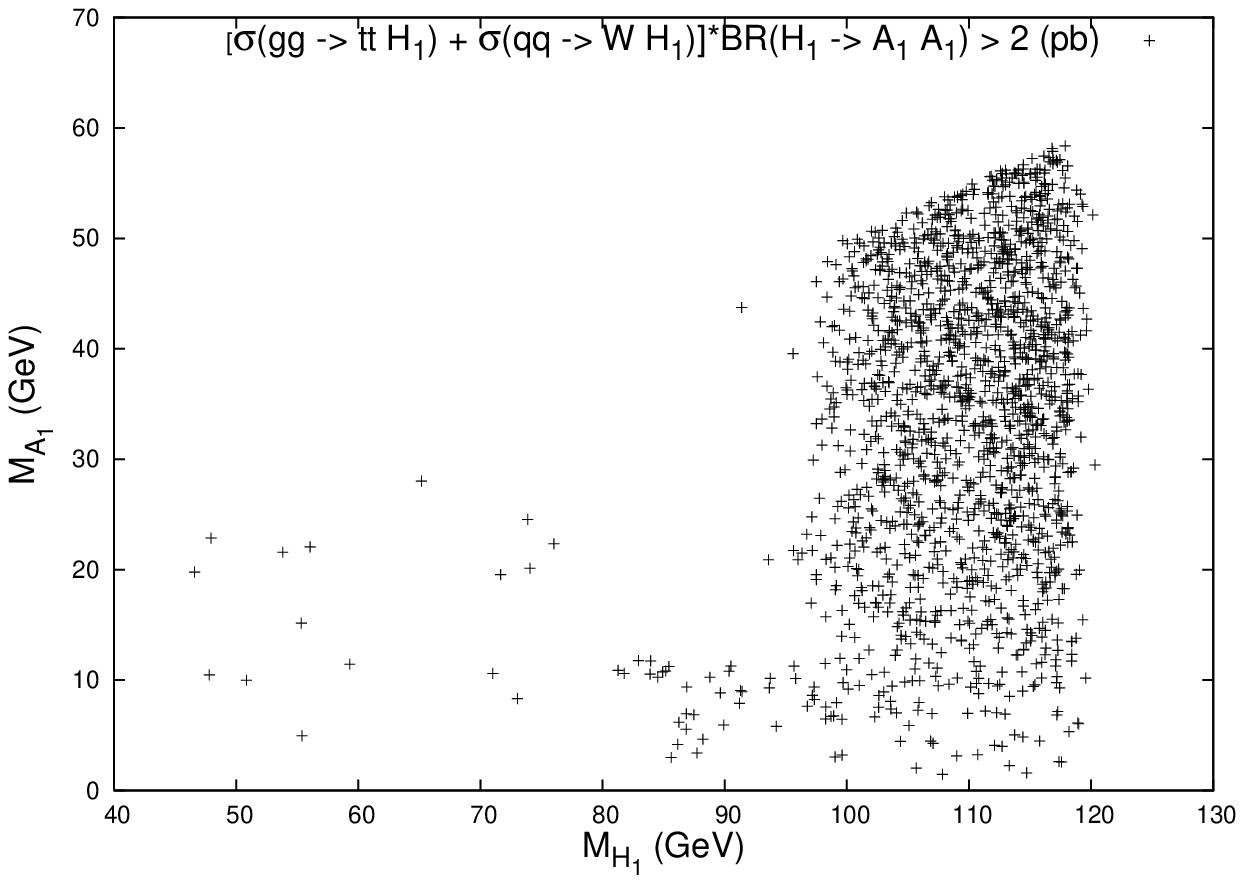}&
\hspace*{1.0truecm}\includegraphics[scale=0.505]{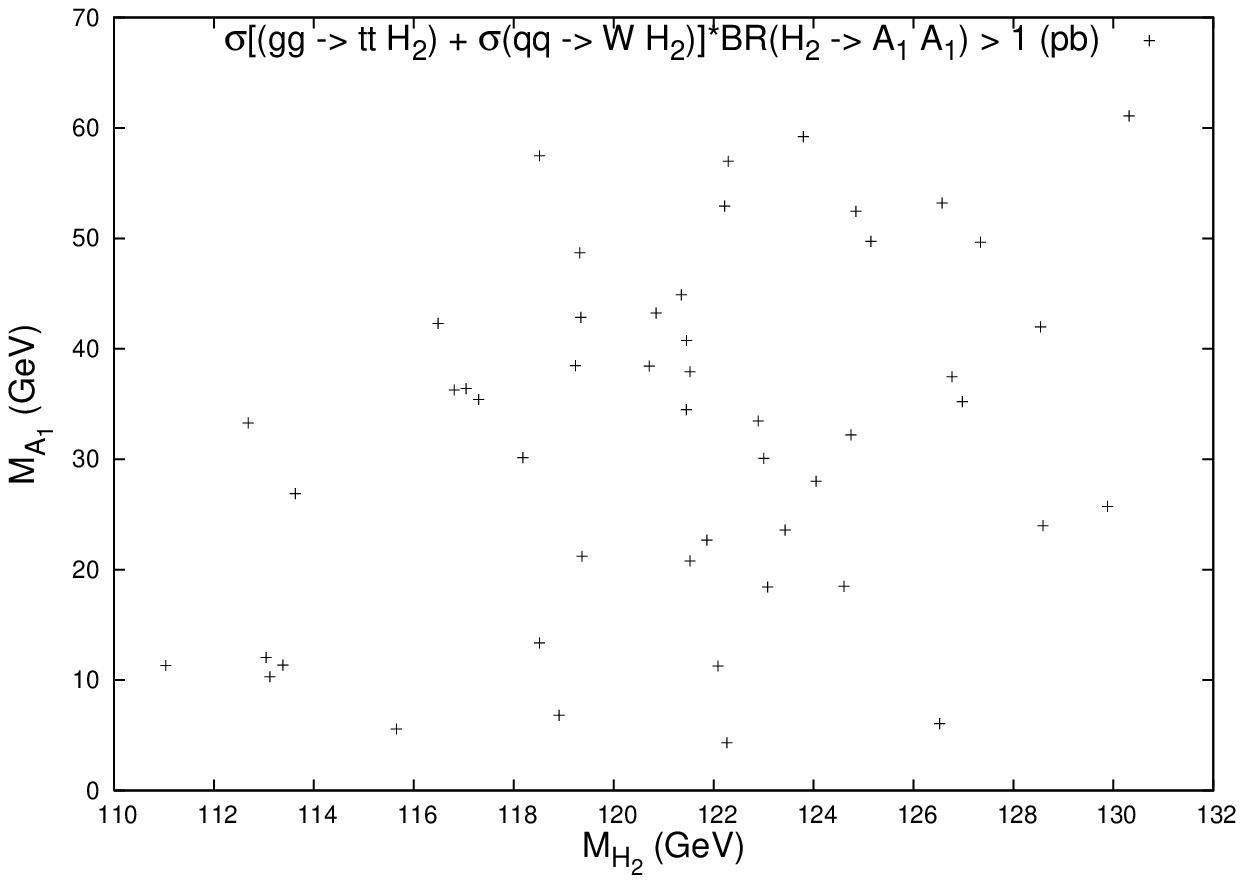}
\end{tabular}
\caption{Distribution of the $H_1$ (left) and $H_2$ (right) masses with respect to that of $A_1$,
when VBF (top) and W-HS + tt-HS (bottom) are potentially visible, i.e., limited to those NMSSM parameter
points for which both cross sections times BRs are larger than 2(1) pb for $H_1(H_2)$.}
\label{fig:mass}
\end{figure}

\end{document}